\documentstyle[11pt,paspconf,psfig,twoside]{article}
\begin{document}
\title{Projected XMM Vision for MCVs }
\author{Cynthia H. James, Graziella Branduardi-Raymont and Mark Cropper}
\affil{Mullard Space Science Laboratory, University College London, Holmbury St Mary, Surrey, UK}
\begin{abstract}
The XMM satellite has a scientific payload of 3  EPIC and 2 RGS instruments
which provide for both imaging and non-dispersive spectroscopy as well
as medium-to-high resolution dispersive spectroscopy, in the soft to
medium X-ray bands. The co-aligned OM adds the facility
for simultaneous optical and UV studies. In this poster paper we
describe the instrumental characteristics of XMM, comparing them to
those of ASCA. We also forecast the expected X-ray spectra of a
selection of both the Polars and Intermediate Polars
 by means of simulations based on ASCA model fits.
\end{abstract}
\section{Introduction}

XMM, carries 3 co-aligned X-ray telescopes with (to date) an
unsurpassed X-ray photon collecting potential. At the focus of two of
these telescopes is an RGS dispersive spectrometer and an EPIC-mos
capable of both non-dispersive spectroscopy and imaging. At the focus
of the third X-ray telescope is an EPIC-pn, again capable of both
non-dispersive spectroscopy and imaging.

There is currently considerable interest in the physical processes
associated with accretion and accretion disks. Magnetic Cataclysmic
Variables (mCVs) provide us with Galactic examples of these processes
in the presence of strong magnetic fields, but progress is presently
limited by the data quality. The ASCA SIS is at present the facility
in the X-ray wavelength band which most closely approaches the
capabilities of XMM. Over the years since its launch in 1993 a number
of papers have included the spectral analysis of several Polar and
Intermediate Polar (IPs).

In this poster paper we make an assessment of the scientific
capabilities of XMM for mCVs. We use the spectral fitting software
package, XSPEC, to simulate data of several mCVs based on the
parameters from the spectral model fitting of ASCA data.

\section{Simulations}

In order to place these simulations in context we firstly make a
comparison between the ASCA SIS and the XMM RGS and EPIC cameras with
respect to the instrument specifications, as given in Table 1, and
their respective photon gathering efficiency, as given in Figure 1.
The models used in these simulations include both thermal
bremssstralung, and optically thin MEKAL plasma (single and
multi-temperature) codes, with constant and partial covering
absorbers. In all cases we have limited the modelling to that which
was used for the ASCA analysis.  The parameters we've used for the
simulations are given in Tables 2, 3 and 4.  The simulated data,
created by folding the models through published response matrices for
each of the XMM X-ray instruments is shown in Figures 2, 3 and 4.

\begin{table}	
\begin{center}\scriptsize
\begin{tabular}{p{45mm}p{15mm}p{15mm}p{15mm}p{15mm}p{15mm}}
Instrument Specification          & ASCA SIS & EPIC-pn  & EPIC-mos & RGS order 1 & RGS order 2     \\
\tableline
energy range      (keV)           & 0.7-10.0 & 0.1-15.0 & 0.1-10.0 & 0.35-2.50   & 0.62-2.50       \\
energy resolution (eV) @ 0.35 keV &          &          & 0.443    &                               \\
energy resolution (eV) @ 0.69 keV &          &          & 1.44     & 0.84                          \\
energy resolution (eV) @ 1.24 kev &          &          & 4.32     & 2.38                          \\
energy resolution (eV) @ 2.00 keV & ~0.18    & 0.08     & 0.08     &             &                 \\
energy resolution (eV) @ 6.00 keV & ~0.20    & 0.13     & 0.08     &             &                 \\
\end{tabular}
\caption{Comparison of relevant ASCA and XMM instrumental specifications.} \label{tbl-1}
\vspace*{20mm}
\begin{tabular}{p{25mm}p{45mm}l}
integration time  & (s)                            &43000                \\
flux (2-10keV)    & erg cm$^{-2}$ s$^{-1}$         & 8.0 x 10$^{-12}$    \\
metal abundance   & Anders \& Grevesse             & 1.76                \\
isothermal plasma & Single temperature Mekal (keV) & 12.0                \\
partial absorbers 
& nH (10$^{22}$ cm$^{-2}$)       & 3.2 x 10$^{-1}$     \\
                  & Covering fraction              & 0.39                \\
\end{tabular}
\caption{The model and parameters values used in the simulation for BL Hyi as specified in Table 2 model 2, Matt et al. (1998).} \label{tbl-1}
\vspace*{20mm}
\begin{tabular}{p{25mm}p{45mm}l}
integration time  & (s)                       & 34000                   \\
flux @ 0.6-10 keV & erg cm$^{-2}$s$^{-1}$     & 1.4 x 10$^{-10}$        \\
isothermal plasma & 4 temperature Mekal(keV)  & 15.4, 3.2, 1.0, 0.54    \\
metal abundance   & Allen 1973                & 1.0                     \\
                  & normalisation $(10^{-2})$ & 3.62, 3.76, 0.72, 0.67  \\
partial absorber  & nH (10$^{22}$ cm$^{-2}$)  & 0.24                    \\
                  & covering fraction         & 0.60                    \\
\end{tabular}
\caption{The model and parameters values used in the simulation for EX Hya as specified in Table 1 spin max, Allan et al. (1998).} \label{tbl-1}

\vspace*{20mm}

\begin{tabular}{p{25mm}p{45mm}l}
integration time    & (s) &37000 \\
flux @ 2-10 keV     & erg cm$^{-2}$ s$^{-1}$                     & 7.2 x 10$^{-11}$           \\
continuum           & bremsstralung temperature (keV)            & 14.0                       \\
metal abundance     & Anders \& Ebihara                          & 1.0                        \\
3 partial absorbers & nH (10$^{22}$ cm$^{-2}$)                   & 9.0 x 10$^{-3}$, 2.0, 20.0 \\
                    & Covering fraction                          & 0.27, 0.22, 0.51           \\
3 emission lines    & Fe (keV)                                   & 6.40, 6.65, 6.93           \\
                    & sigma (keV)                                & 0.06, 0.04, 0.06           \\
                    & total photons cm$^{-2}$ s$^{-1}$ 10$^{-4}$ & 1.163, 1.17, 1.131         \\
\end{tabular}
\caption{The model and parameters values used in the simulation for AM Her as specified spin maximum, Ishida et al. (1997).} \label{tbl-1}
\end{center}
\end{table}
\begin{figure}
\begin{minipage}[t]{65mm}
\psfig{file=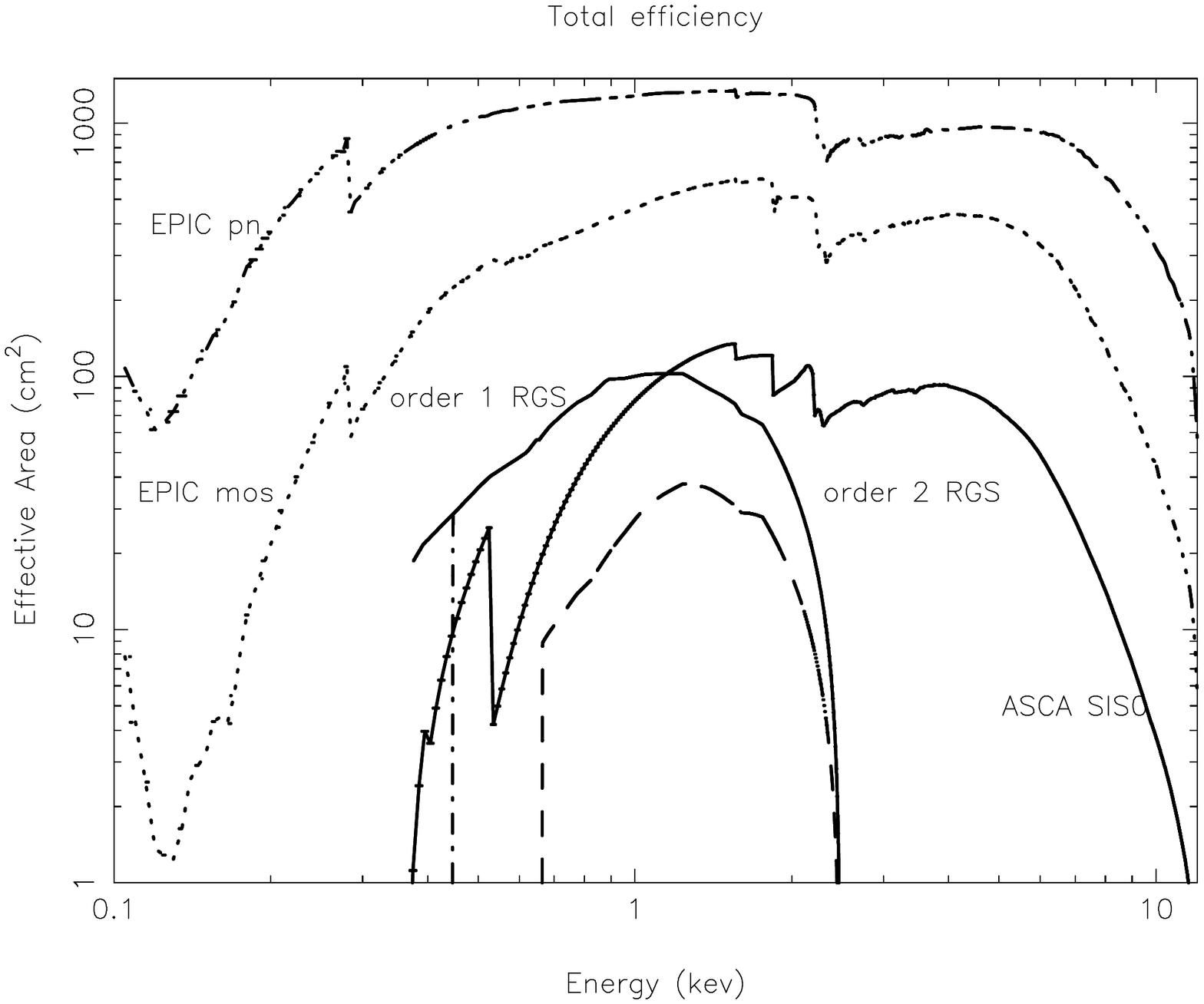,width=61mm,angle=270,rwidth=20mm}
\caption{Comparison of effective areas between ASCA and the XMM X-ray
  instruments.}
\end{minipage}
\hspace*{5mm}
\begin{minipage}[t]{65mm}
\psfig{file=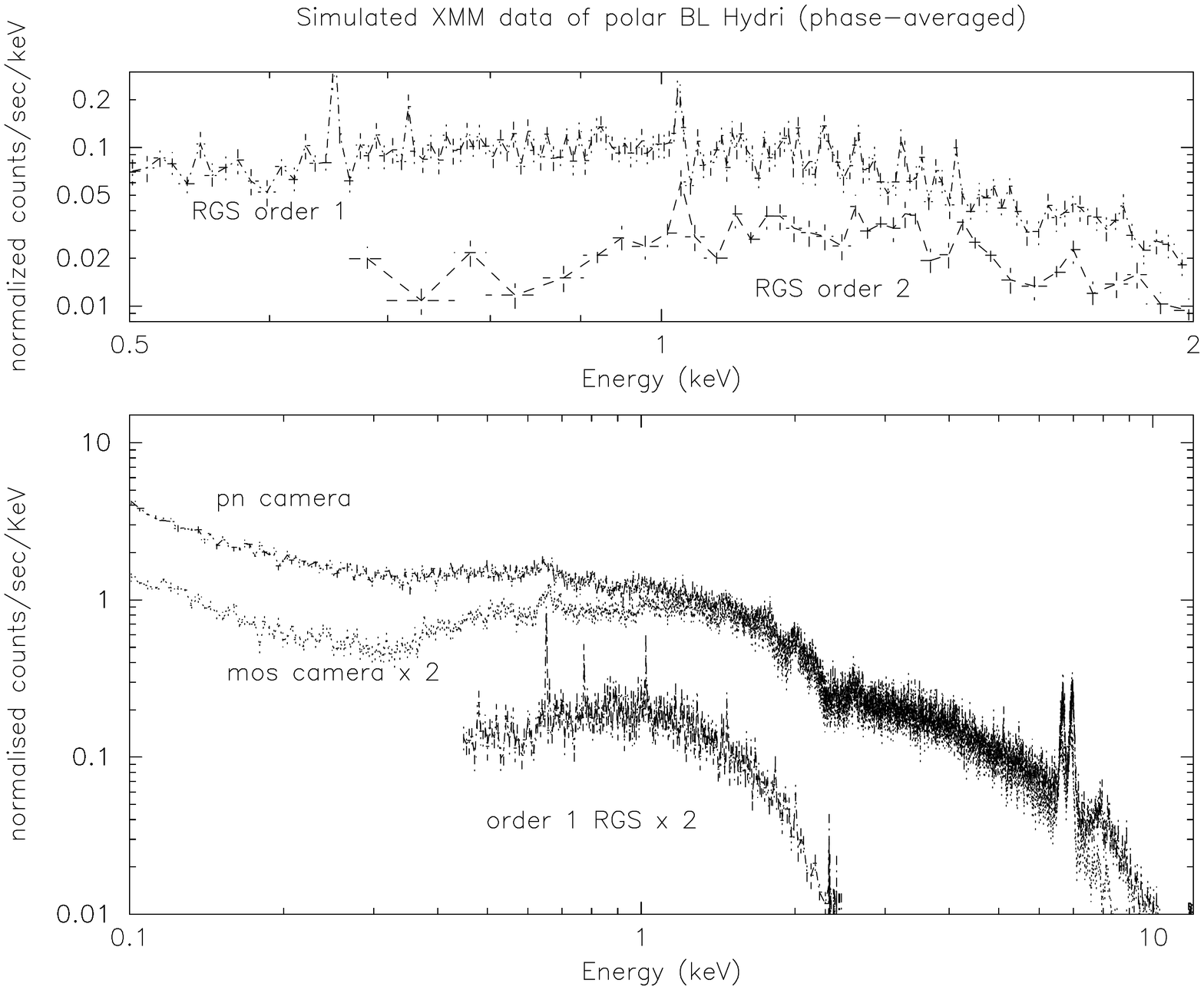,width=65mm,angle=270,rwidth=20mm}
\caption{Simulated XMM data of the Polar BL Hydri (phase averaged), using the 
model parameters as given in Table 2.  
Top: orders 1 and 2 spectra from a  single RGS. 
Bottom: spectra from the two RGS, the two EPIC-mos  
and single EPIC-pn cameras.}
\end{minipage}
\end{figure}
\begin{figure}
\begin{minipage}[t]{65mm}
\psfig{file=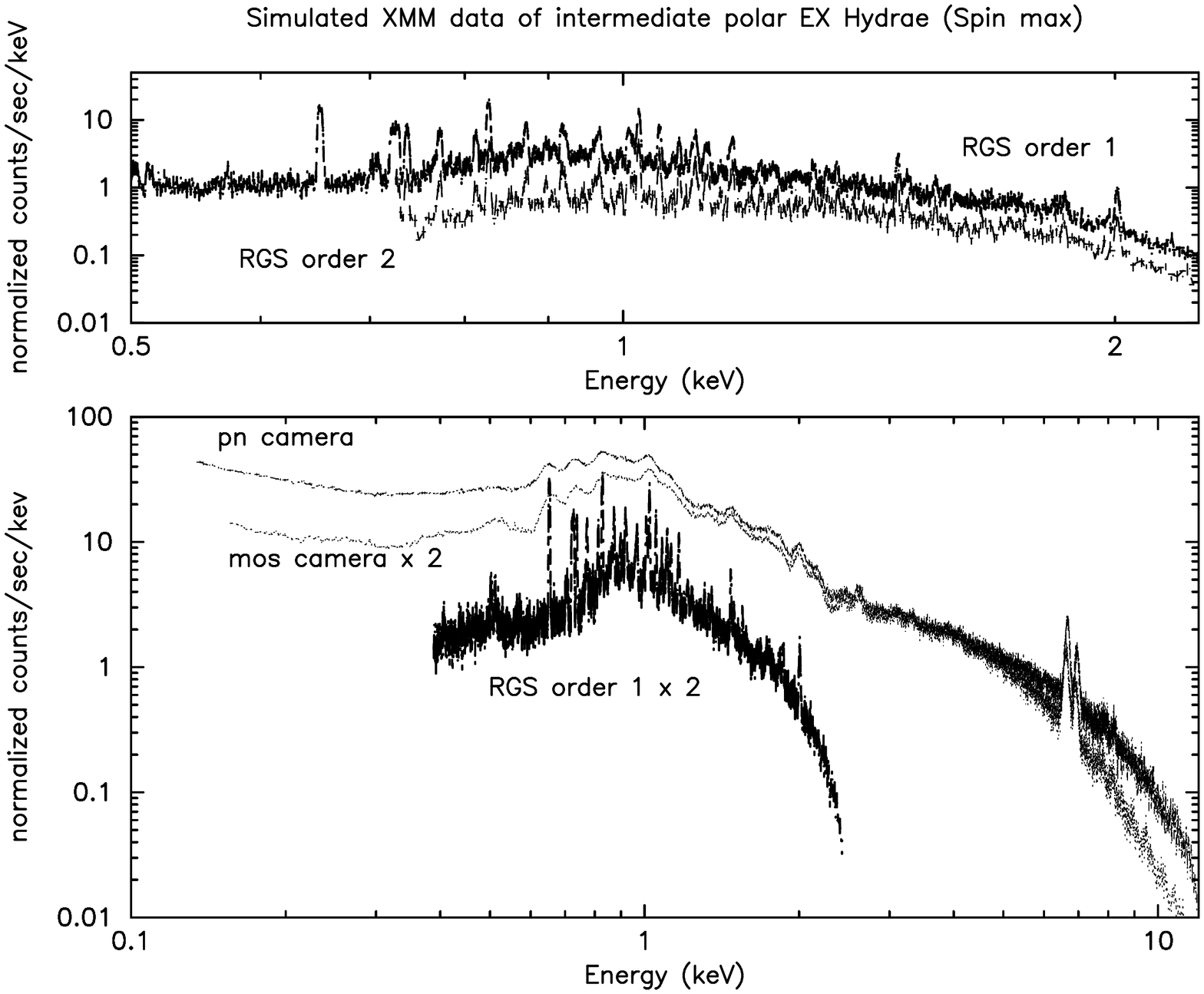,width=65mm,angle=270}
\caption{Simulated XMM data for IP EX Hya (spin maximum) using the model parameters as given in Table 3. 
Top: orders 1 and 2 spectra from a  single RGS.  
Bottom: spectra from the two RGS, the two EPIC-mos  and single EPIC-pn cameras.}
\end{minipage}
\hspace*{-1mm}
\begin{minipage}[t]{65mm}
\psfig{file=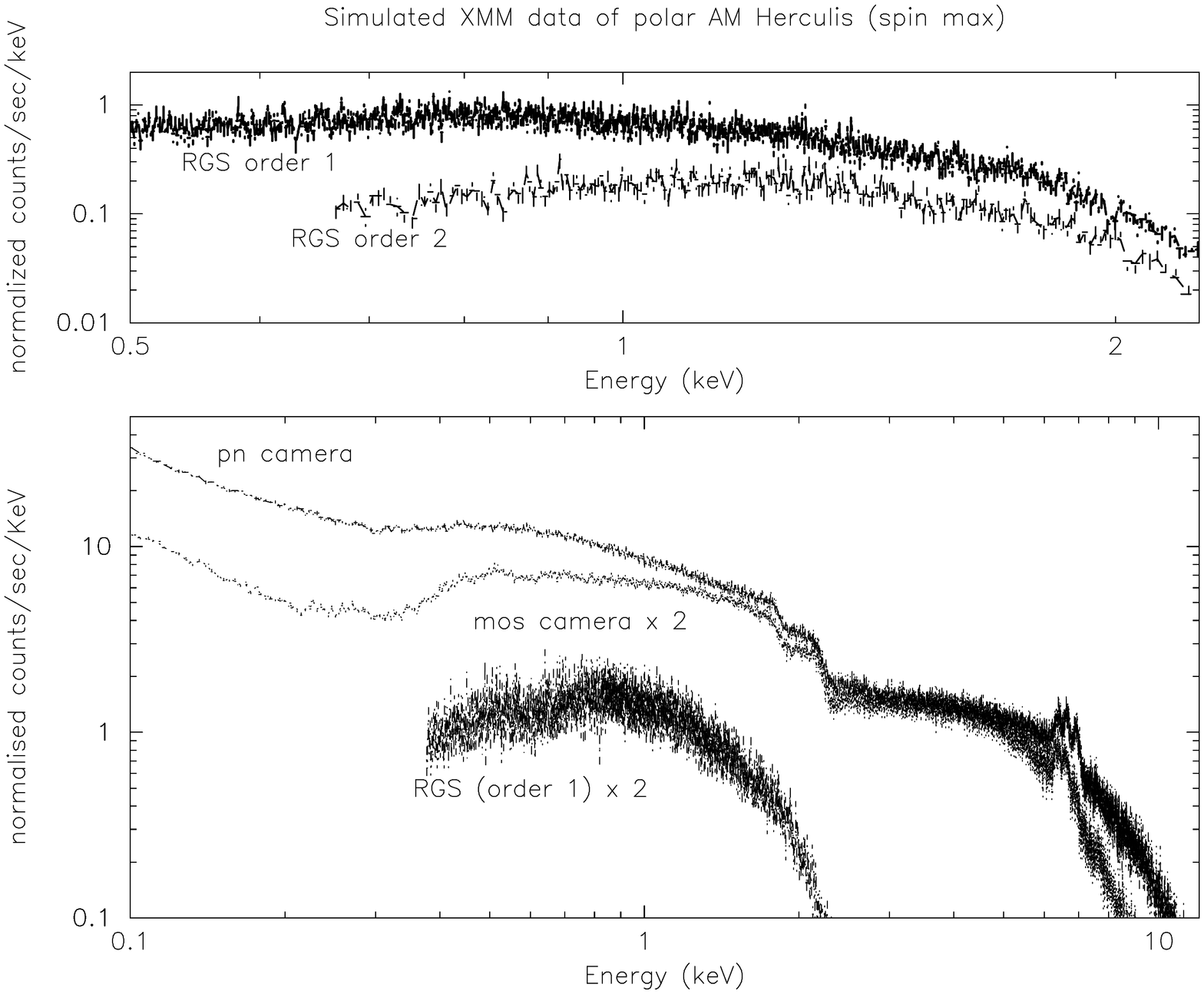,width=65mm,angle=270}
\caption{Simulated XMM data for polar AM Her at spin maximum using the model parameters as given in Table 4. 
Top: orders 1 and 2 spectra from a  single RGS.  
Bottom: spectra from the two RGS, the two EPIC-mos  and single EPIC-pn cameras.}
\end{minipage}
\end{figure}
\section{Discussion}

Probably the most dramatic improvement in data quality is with respect
to the effective area. This can be appreciated by comparing the
normalised counts/sec/keV given by simulated EPIC camera results from
BL Hyi data, Figure 2, EX Hya data, Figure 3 and AM Her data, Figure 4
with their respective results from ASCA data, Matt et al. (1998),
Allan et al. (1998) and Ishida et al. (1997).  Even in the case of the
RGS the combined output of the first order spectra equates to that
obtained from the ASCA data over its relevant wavelength range.

Figures 2 -- 4 clearly demonstrate the improvement in energy
resolution by the number of resolvable emission lines present in the
simulated data.  This is clear from a comparison of simulated spectra
of BL Hyi data, Figure 2, and EX Hya data , Figure 3, with their ASCA
spectra in Matt et al. (1998) and Allan et al. (1998) respectively.
The underlying model fitted for AM Her is a bremsstralung continuum,
hence this effect is less apparent.  Another illustration is the
comparison of the simulated spectra of EX Hya and BL Hyi with that of
AM Her.  All three systems show Fe emission lines in the 6.0 -- 7.0
keV range.  However in the latter their presence is directly
attributable to the added Gaussians based on ASCA fitting parameters:
these are much broader than in the first pair, where their presence is
due to the metal abundance parameter in the MEKAL code model.
  
Finally, the resolution of the RGS first order is probably
insufficient to resolve velocity shifts of the accretion flows in
these systems. However, where the level of flux is relatively high, as
in the case of EX Hya, then the higher resolution of the RGS 2nd order
point spread function may be sufficient.

\end{document}